\documentclass[epj]{webofc}
\usepackage[utf8]{inputenc}
\usepackage[varg]{txfonts}   % Web of Conferences font
\usepackage{booktabs}
\usepackage{xcolor}

\usepackage{bbm}
\usepackage[utf8]{inputenc}
\usepackage{graphicx}
\usepackage{graphics}
\usepackage{ae}
\usepackage{amsmath}
\usepackage{amssymb}
\usepackage{slashed}	
\usepackage{lmodern}
\usepackage{framed}
\usepackage{colortbl}
\usepackage{cancel}
\usepackage{arydshln}
\usepackage{wasysym}
\usepackage{multido}
\usepackage{subfigure}

\usepackage[dvips]{epsfig}
\usepackage{bm}
\usepackage{bbm}
\usepackage{verbatim}
\usepackage{psfrag}
\usepackage{slashed}
\usepackage[normalem]{ulem}

\usepackage{amsmath}
\usepackage{amssymb}
\usepackage{wasysym}
\usepackage{multirow}
\usepackage{youngtab}

\usepackage{tcolorbox}% http://ctan.org/pkg/tcolorbox

\makeatletter
\newcommand{\raisemath}[1]{\mathpalette{\raisem@th{#1}}}
\newcommand{\raisem@th}[3]{\raisebox{#1}{$#2#3$}}
\makeatother

\definecolor{darkred}{rgb}{0.4,0.0,0.0}
\definecolor{darkgreen}{rgb}{0.0,0.4,0.0}
\definecolor{darkblue}{rgb}{0.0,0.0,0.4}
\usepackage[bookmarks,linktocpage,colorlinks,
    linkcolor = darkred,
    urlcolor  = darkblue,
    citecolor = darkgreen]{hyperref}
\usepackage{subfigure}
\wocname{EPJ Web of Conferences}
\woctitle{Lattice2017}
\input{newcommand.in}

\newcommand{\bkappa}{\bar{\kappa}}
\newcommand{\Bm}{\bar{m}}

\newcommand{\hmu}{{\hat{\mu}}}
\newcommand{\hnu}{{\hat{\nu}}}

\newcommand{\Nt}{{N_{\tau}}}

\newcommand{\M}{\mathcal{M}}
\newcommand{\Md}{\mathcal{M}^\dagger}

\newcommand{\tr}{{\rm tr}}

\newcommand{\ijkl}{_{\bm{i\,\raisemath{3pt}{j},k\,\raisemath{3pt}{l}}}}
\newcommand{\jilk}{_{\bm{j\,\raisemath{3pt}{i},l\,\raisemath{3pt}{k}}}}

\renewcommand{\ij}[1]{{}_{i_{#1}}^{\;\;j_{#1}}}
\newcommand{\ji}[1]{{}_{j_{#1}}^{\;\;i_{#1}}}
\newcommand{\kl}[1]{{}_{k_{#1}}^{\;\;l_{#1}}}
\newcommand{\lk}[1]{{}_{l_{#1}}^{\;\;k_{#1}}}

\newcommand{\IJZZ}{_{\bm{{}_i{}^j}}}

\newcommand{\I}{\mathcal{I}}
\newcommand{\J}{\mathcal{J}}
\newcommand{\K}{\mathcal{K}}

\newcommand{\Wg}[1]{{{\rm Wg}^{#1}}}

\newcommand{\Ud}{U^\dagger}

\newcommand{\sgn}{{\rm sgn}}

\begin{document}
%%%%%%%%%%%%%%%%%%%%%%%%%%%%%%%%%%%%%%%%%%%%%%%%%%%%%%%%%%%%%%%%%%%%%%%%%%%%%
%
\selectlanguage{english}
%----------------------------------------------------------------------------
\title{%
Dual Formulation and Phase Diagram of Lattice QCD\\in the Strong Coupling Regime
}
%----------------------------------------------------------------------------
\author{%
\firstname{Giuseppe} \lastname{Gagliardi}\inst{1} \and
\firstname{Jangho} \lastname{Kim}\inst{1,2} \thanks{Speaker, \email{jangho@physik.uni-bielefeld.de}}\and 
\firstname{Wolfgang} \lastname{Unger}\inst{1}\thanks{Speaker, \email{wunger@physik.uni-bielefeld.de}} 
% etc.
}
%----------------------------------------------------------------------------
\institute{%
Fakult\"at f\"ur Physik, Universit\"at Bielefeld, D-33615 Bielefeld, Germany \and
National Superconducting Cyclotron Laboratory and Department of Physics and Astronomy,\\Michigan State University, East Lansing, Michigan 48824, USA
}
%----------------------------------------------------------------------------
\abstract{%
We present the computation of invariants that arise in the strong coupling expansion of lattice QCD. 
These invariants  are needed for Monte Carlo simulations of Lattice QCD with staggered fermions in a dual,
color singlet representation. This formulation is in particular useful to tame the finite density sign problem.
The gauge integrals in this limiting case $\beta\rightarrow 0$ are well known, but the gauge integrals needed to study the gauge corrections 
are more involved. We discuss a method to evaluate such integrals.

The phase boundary of lattice QCD for staggered fermions in the $\mu_B-T$ plane has been established in the strong coupling limit.
We present numerical simulations away from the strong coupling limit, taking into account the higher order gauge corrections via plaquette occupation numbers.
This allows to study the nuclear and chiral transition as a function of $\beta$.
}
%----------------------------------------------------------------------------
\maketitle
%----------------------------------------------------------------------------
\section{Introduction}\label{intro}

The finite baryon density sign problem in lattice QCD hinders a direct evaluation of the phase structure of QCD in the $\mu_B-T$ plane. In particular, 
the existence of a critical end-point (CEP) that is sought for in heavy ion collision experiments at RHIC and LHC could not be established yet via lattice simulations.
Although the well established methods for small $\mu_B/T$, such as Taylor expansion, reweighting and analytic continuation from imaginary chemical potential can in principle 
make statements about the existence of the CEP, it is likely that the CEP, if it exists, has a quite large $\mu_B^{crit}$, 
such that it is not within reach with the aforementioned methods.

In recent years, many alternative methods have been proposed and tested to circumvent the finite density sign problem. Most notably, 
the complex Lagenvin method together with gauge cooling could address full QCD in the deconfined phase \cite{Sexty:2013ica,Aarts:2013uxa}.
Another method based on complexified QCD, the Lefschetz thimbles,
are currently applied to QCD-inspired models with few degrees of freedom, but the method is far from being applicable to full lattice QCD \cite{Alexandru:2015sua,Schmidt:2017gvu,DiRenzo:2017igr}.\\

A promising alternative strategy is to change the degrees of freedom of the original partition function. Since the sign problem is representation dependent, 
it may be possible to find a different set of variables that are closer to the true eigenstates of the Hamiltonian. 
Finding such a basis would reduce the sign problem significantly, or even solve it. 
Changing the degrees of freedom can be for example obtained by a Hubbard-Stratonovich transformation \cite{Vairinhos:2014uxa}, or by introducing auxiliary fields.
Another way is to integrate out some of the degrees of freedom to obtain a ``dual'' representation in terms of world lines.
This strategy has been successfully applied to address sign problems in models with an abelian gauge group (such as the massless Schwinger model
\cite{Gattringer:2015nea}, and the gauge-Higgs models \cite{Mercado:2013yta})	
It is however quite non-trivial to find a dual representation for non-abelian gauge groups. 
A recent attempt is to decompose the non-abelian components into abelian ``color cycles'' \cite{Gattringer:2016lml}. 

Our attempt to perform Monte Carlo simulations on the QCD phase diagram is based on the strong coupling expansion. The starting point is the well-established 
partition function of staggered fermions in the strong coupling limit. Here, the phase diagram is well established.
We then propose a dual representation in terms of world lines and world sheets that incorporates some contributions of the gauge action. For small $\beta$, 
we are able to determine the phase boundary between the chirally broken and chirally restored phase. The leading order correction has been addressed via reweighting
from the strong coupling ensemble to $\beta>0$ in \cite{deForcrand:2014tha}. 
We go beyond this scope by directly sampling the partition function including next to leading order gauge corrections.

%----------------------------------------------------------------------------
\section{Link Integration}\label{sec-1}

\subsection{Lattice Action and Partition Function}\label{sec-1-1}

We consider the standard lattice action for staggered fermions (no rooting, no improvement) together with the Wilson gauge action:
\begin{align}
S_{\rm F}&= \sum_{x}\left(\sum_{\mu} \gamma^{\delta_{\mu 0}}
\eta_\nu(x)\left(e^{a_t \mu\delta_{\mu 0}} \bar{\chi}_x U_\mu(x) \chi_{x+\hmu} - e^{-{a_t\mu}\delta_{\mu 0}}  \bchi_{x+\hmu} U_\mu^\dagger(x) \chi_x \right)
 + 2am_q\bar{\chi}_x\chi_x\right),\\
S_{\rm G}&= \frac{\beta}{2\Nc	} \sum_{P=(x,\mu<\nu)} \tr[U_P+U_P^\dagger],\qquad U_P=U_\mu(x)U_\nu(x+\hmu)U_\mu(x+\hnu)^\dagger U_\nu(x)^\dagger,
  \end{align}
with $a_t \mu=\frac{1}{\Nc}a_t \mu_B$ the quark chemical potential. The only modification is that we introduced a bare anisotropy $\gamma$, favoring temporal fermion hoppings over spatial fermion hoppings, 
giving rise to an anisotropy of the lattice spacings $\frac{a}{a_t}=\xi(\gamma)$. This will allow us later to vary the temperature continuously in the strong coupling regime.
  
The standard approach for lattice simulations is to integrate out the Grassmann-valued staggered fermions $\chi$ and $\bchi$ to obtain the fermion determinant. However, 
the fermion determinant becomes complex for finite quark chemical potential, resulting in the finite density sign problem. 
Our strategy is to expand the action $S=S_F+S_G$ both in the fermion hoppings and in $\beta=\frac{2\Nc}{g^2}$.
Then we exchange the order of integration, i.e.~integrate out the link variables analytically first, and afterwards the Grassmann variables.
The remaining degrees of freedom will be color singlets on the links, and the plaquette occupation numbers $n_P$ (from the moments of the fundamental plaquettes $\tr[U_P]^{n_P}$)
and $\bar{n}_P$ (from the moments of the anti-fundamental plaquettes $\tr[U_P^\dagger]^{\bar{n}_P}$).\\

\begin{figure}[thb] 
  \centering
  \includegraphics[width=0.47\textwidth,clip]{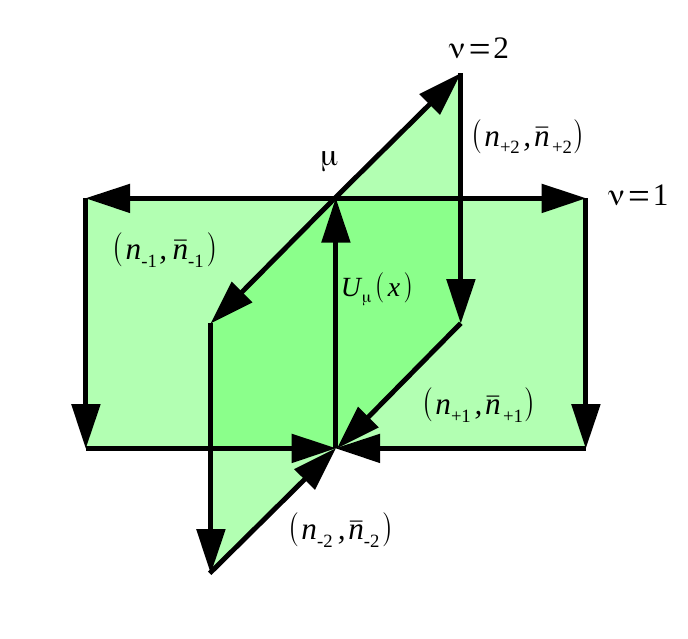}
  \caption{Staples and corresponding plaquette occupation numbers $(n_P,\bar{n}_P)$ for directions perpendicular to the link $U_\mu(x)$ to be integrated out: 
  the moment of $U_\mu(x)$ from the moments of the Wilson gauge action is $m=\sum\limits_{\nu\perp\mu}  n_{+\nu}+\bar{n}_{-\nu}$ and 
  $\bar{m}=\sum\limits_{\nu\perp\mu}  n_{-\nu}+\bar{n}_{+\nu}$, see Eq.~(\ref{MConstr}).
}
\label{fig-1-1}
\end{figure}

The fermions can be gathered into matrices 
\begin{align}
\Md\ji{}&= \eta_\mu(x)\bchi{}^i_x{\chi}_{x+\hmu,j},& \M\lk{} &= - \eta_\mu(x)\bchi{}^k_{x+\hmu} \chi{}_{x,l} = \eta_\mu(x) \chi_{x,l}\bchi{}^k_{x+\hmu}. 
\end{align}
All elementary plaquettes $P$ from the expansion of $S_{\rm G}$ that share a given link $U_\mu(x)$ 
need to be taken into account when integrating out the link $U\equiv U_\mu(x)$: 
\begin{align}
P_U&=\{P\; |\; U\in U_P\}=P_U^+\cup P_U^-. 
\end{align}
with $P_+$ the subsets of plaquettes in forward and $P_-$ in backward direction, as illustrated in Fig.~\ref{fig-1-1}.
Hence the one-link integral over gauge group $G=\SU(\Nc), U(\Nc)$ that we will consider has the fermion matrices $\M$, $\Md$ and the set of staples $S_P$ with $U_P= U_\mu(x) S_P$ as external sources:
\begin{align}
\J%^{a,b}\ijkl
&(\M,\Md,\{S_P,S_P^\dagger\})=\int_{G} dU\, e^{\tr[\Md U]+ \tr[\M\Ud]}e^{\frac{\beta}{2\Nc} {\sum\limits_{P\supset U}}(\tr[US_P]+ \tr[\Ud S^\dagger_P])}\nonumber\\
&=\int_{G} dU\sum_{\kappa,\bkappa} \frac{\tr[\Md U]^{\kappa} \tr[\M\Ud])^{\bkappa} }{\kappa!\bkappa!} \prod_{P\supset U}\sum_{n_P,\bar{n}_P}
\left(\frac{\beta}{2\Nc}\right)^{n_P+\bar{n}_P}
\frac{\tr[US_P]^{n_P} \tr[\Ud S_P^\dagger]^{\bar{n}_P}}{n_P!\bar{n}_P!} \nonumber\\
&=\sum_{\kappa,\bkappa} \prod_{P\supset U}\sum_{n_P,\bar{n}_P}\frac{1}{\kappa!\bkappa!}\frac{
\left(\frac{\beta}{2\Nc}\right)
^{n_P+\bar{n}_P}}{n_P!\bar{n}_P!}\int_{G} dU\tr[\Md U]^{\kappa} \tr[\M\Ud]^{\bkappa} 
 \tr[US_P]^{n_P} \tr[\Ud S^\dagger_P]^{\bar{n}_P}\nonumber\\
 &= \sum_{m,\bar{m}} C(\beta,\{S_P,S^\dagger_P\})^{m,\Bm}\jilk\;\sum_{\kappa,\bkappa}\frac{1}{\kappa!\bkappa!}\K^{m,\Bm}\ijkl(\M,\Md),\qquad \tr[US_P]=\sum_{i,j=1}^\Nc U\ij{}S_P\ji{}
\label{JInt}
 \end{align}
where we expand in the forward hoppings $\kappa$, backward hoppings $\bkappa$, and plaquette and anti-plaquette occupation numbers $n_P$, $\bar{n}_P$.
In the last line, we have decomposed the traces to separate the staples from the gauge link and summation over the set of indices $\bm{i}$, $\bm{j}$, $\bm{k}$ ,$\bm{l}$ is implied.
It is the tensor $C(\beta,\{S_P,S^\dagger_P\})\jilk$ which leads to non-local color contractions and can be related to the set of plaquette occupation numbers $\{n_P,\bar{n}_P\}$ 
when contracting the $m$ open color indices from $U$ and $\bar{m}$ open color indices from $\Ud$ with the one-link integrals from the neighbor links:
\begin{align}
m&=\sum_{P\in P^+_U} n_{P}+\sum_{P\in P^-_U}\bar{n}_{P},&
\Bm&=\sum_{P\in P^+_U} \bar{n}_{P}+\sum_{P\in P^-_U}n_{P}.
\label{MConstr}
\end{align}
The remaining integral can be related to integrals over the link matrices only \cite{Creutz1978}:
\begin{align}
\I^{a,b}\ijkl&=\int_G dU \prod_{\alpha=1}^a U\ij{\alpha} \prod_{\beta=1}^b (\Ud)\kl{\beta}
,\qquad 
\begin{array}{ll}
\bm{i}=i_1,\ldots i_a & \bm{k}=k_1,\ldots k_b\\
\bm{j}=j_1,\ldots j_a & \bm{l}=l_1,\ldots l_b
\end{array}
\label{IInt}\\
 \K^{m,\Bm}\ijkl
(\M,\Md)&=\int_{G} dU\tr[\Md U]^{\kappa} \tr[\M\Ud]^{\bkappa} 
 \prod_{\alpha=1}^m U\ij{\alpha} 
 \prod_{\beta=1}^{\Bm} {\Ud}\kl{\beta}\nonumber\\
 &= \sum_{\{i_\alpha,j_\alpha,k_\beta,l_\beta\}}
\left( \prod_{\alpha=1}^{\kappa}\prod_{\beta=1}^{\bkappa}\M\ji{\alpha}\Md\lk{\beta}\right)
{\I^{\kappa+m,\bkappa+\Bm}\ijkl}.
\end{align}
Here, $a=\kappa+m$ and $b=\bkappa+\Bm$ is the number of $U$-matrix and $\Ud$-matrix elements.
In this one-link integral, only the color indices from the quark matrices will be contracted.
The contraction of the remaining indices can in general not be carried out easily, 
however in certain cases link integration on the complete lattice will be possible to give rise to a color singlet partition function:
\begin{align}
Z(\beta)=\sum_{\mathcal{G}=\{n_P,\bar{n}_P,\kappa,\bkappa\}} w(\mathcal{G})\prod_P\left(\frac{\beta}{2\Nc}\right)^{n_P+\bar{n}_P},
\end{align}
where the admissible graphs $\mathcal{G}$ are such that they fulfill the constraint
\begin{align}
\kappa-\bar{\kappa}+m-\bar{m}=\left\{
\begin{array}{lll}
0 & \text{for} &\U(\Nc)\\
0 \mod \Nc & \text{for} &\SU(\Nc)
\end{array}
\right.. 
\end{align}

\subsection{Link Integration in the Strong Coupling Limit}\label{sec-1-2}

For $\beta=0$, link integration factorizes: 
\begin{align}
Z_0(am_q,a_t\mu,\gamma)=&\prod_x \int d\chi_x d\bchi_x e^{2am_q\bchi_x\chi_x}\nonumber\\
&\prod_\mu  \int dU_\mu(x) \left(
e^{\gamma^{\delta_{\mu 0}}\eta_\nu(x)e^{a_t \mu\delta_{\mu 0}} 
(\bar{\chi}_x U_\mu(x) \chi_{x+\hmu}-\bar{\chi}_{x+\hmu} U^\dagger_\mu(x) \chi_{x})
}
\right).	
\end{align}
The corresponding one-link integrals $\K_0$ will not depend on any external gauge links:
\begin{align}
 \K_0%^{a,b}\ijkl
(\M,\Md)&=\int_{G} dU\tr[\Md U]^{\kappa} \tr[\M\Ud]^{\bkappa}
 &=\hspace{-4mm} \sum_{\{i_\alpha,j_\alpha,k_\beta,l_\beta\}}
\left(\prod_{\alpha=1}^{\kappa}\prod_{\beta=1}^{\bkappa}\M\ji{\alpha}\Md\lk{\beta}\right)
{\I^{\kappa,\bkappa}\ijkl}.
\end{align}
Hence, link integration can be carried out analytically. Only a finite number of integrals have to be evaluated due to the Grassmann nature of the fermions: 
since they come in $\Nc$ colors, $0\leq \kappa_,\bkappa\leq \Nc$. Moreover, integral Eq.~(\ref{IInt}) will only be non-zero if $\kappa-\bkappa=q\Nc$ with $q=0,\pm 1$ (see next section).
The corresponding result for Eq.~(\ref{JInt}) was first addressed in \cite{Rossi1984} when deriving the strong coupling partition function for $\Nf=1$:
\begin{align}
\J_0(\M,\Md)&=
\int_G dU e^{\tr[U\Md+\M\Ud]}\nonumber\\
%&=\sum_{k=0}^\Nc\left\{\frac{(\Nc-k)!}{\Nc!k!}\left((\eta_\nu(x))^2\bchi_x\chi_x\bchi_{x+\hmu}\chi_{x+\hmu}\right)^k\right\}
%\qquad + \frac{q}{\Nc!}\left\{(\rho_\nu(x)\bchi_x\chi_{x+\hnu})^\Nc+(-\rho_\nu(x)\bchi_{x+\hnu}\chi_x)^\Nc\right\}
&=\sum_{k=0}^\Nc\frac{(\Nc-k)!}{\Nc!k!}\left(M_x M_{x+\hmu}\right)^k
+ \frac{q}{\Nc!}\left\{(\rho_\nu(x)^\Nc \bar{B}_xB_{x+\hmu})+(-\rho_\nu(x))^\Nc \bar{B}_{x+\hmu} B_x)\right\}\nonumber\\
\text{with}\quad q&=\left\{
\begin{array}{lll}
0 & \text{for} & G=\U(\Nc)\\
1 & \text{for} & G=\SU(\Nc)\\
\end{array}
\right. \quad \text{and}\quad  \rho_\nu(x)=\eta_{\nu}(x)\left\{
\begin{array}{cc}
e^{\pm a_t\mu} &\nu=0\\
1 & \text{else}\\
\end{array}
\right..
\label{JSC}
\end{align}
Here, $M_x=\bar{\chi}_x\chi_x$ are the mesonic and $B_x=\frac{1}{\Nc!}\epsilon_{i_1\ldots i_\Nc}\chi_{x,i_1}\ldots \chi_{x,i_\Nc}$ are the baryonic degrees of freedom.
After the final Grassmann integration, where also the expansion of $e^{2am_q \bchi\chi}$ enters, the partition function is exactly rewritten in terms of integer variables:
\begin{align}
Z_0(am_q,a_t\mu_B,\gamma)= \sum\limits_{\{k,n,\ell\}}\prod\limits_{b=(x,\mu)}\frac{(\Nc-k_b)!}{\Nc!k_b!}
\gamma^{2 k_b\delta_{\mu 0}}\prod\limits_{x}\frac{\Nc!}{n_x!}(2am_q)^{n_x} \prod\limits_\ell w(\ell,a_t\mu_B)
\end{align}
where $k_b\in \{0,\ldots, \Nc\}$ are the so-called dimers, i.e.~multiplicities of bonds $b$ that represent meson hoppings, 
$n_x\in \{0,\ldots, \Nc\}$ are the so-called monomers and represent $\bchi_x\chi_x$ not being part of dimers, and the baryon world lines $\ell$ form oriented self-avoiding loops, with loop weight
\begin{align}
w(\ell,a_t\mu_B)&=\frac{1}{\Nc!}\sigma(\ell)\gamma^{\Nc N_{0,\ell}} e^{ \Nt a_t \mu_B r_\ell}, & \sigma(\ell)&=(-1)^{1+r_\ell+N_{-,\ell}}\prod_{(x,\mu)\in \ell}\eta_\mu(x).
\end{align}
Here, $N_{0,\ell}$ is the number of temporal baryon segments on $\ell$.
The sign $\sigma(\ell)$ of a baryon loop $\ell$ is due to geometry: number of backward directions $N_{-,\ell}$, winding number $r_\ell$ and staggered phases along the loop. 
The sign of a configuration is the product of the signs of all baryonic loops. 
The sign problem of sampling this partition function is however very mild for any value of the chemical potential, because the baryons are heavy and hence tend to have 
simple geometries which contribute with positive signs.

\subsection{Weingarten Functions}\label{sec-1-3}

In order to obtain the partition function away from the strong coupling limit, we will make use of Weingarten functions \cite{Weingarten1978,Collins2003}.
This is particularly useful since when some of the link matrices emerge from the Wilson gauge action, we also need contributions to Eq.~(\ref{IInt}) for $a>0$ and $b>0$.
For $n\equiv a=b$, the result is expressed via permutations $\sigma,\tau\in S_n$ on the color indices that go into $2n$ Kronecker deltas, 
and are multiplied by the Weingarten functions, 
which sums over all irreducible representations (irreps) $\lambda$ of $\SU(\Nc)$ that are tensors of $n$ fundamental irreps:
\begin{align}
\I^{n,n}\ijkl&=\sum_{\sigma,\tau\in S_n} 
\prod_{r=1}^n
\left(
\delta_{i_{{\sigma(r)}}}^{\;l_{r}}
\delta_{k_{r}}^{\;j_{\tau(r)}}
\right)
\Wg{n,\Nc}([\sigma\circ \tau^{-1}]),\\
\Wg{n,N}(\rho)&=\frac{1}{(n!)^2}\sum_{\lambda\vdash n,l(\lambda)\leq\Nc} \frac{(f^\lambda)^2}{D_\lambda(N)}\chi_\lambda^\rho
%,\quad \tWg{n,N}(\rho)=\frac{1}{(n!)^2}\sum_{\lambda\vdash n} \frac{(f^\lambda)^2}{D_\lambda(N)}\chi_\lambda^\rho
\label{Weingarten},
\end{align}
with $D_\lambda(N)$ the dimension of the irrep $\lambda$ of $\U(N)$ and $f^\lambda$ the dimension of the irrep $\lambda$ of $S_n$.
The irreps of both the unitary and the symmetric groups are labeled by integer partitions 
\begin{align}
\lambda&\vdash n,& \lambda&=(\lambda_1,\ldots \lambda_{l(\lambda)}),& n&=\sum_{i=1}^{l(\lambda)} \lambda_i, &\lambda_i\geq \lambda_{i+1},
\end{align}
and due to the finite number of available color indices, 
the corresponding partitions have a finite number of parts, $l(\lambda)\leq N$.
The Weingarten functions contain the character $\chi_\lambda^\rho$ of the symmetric group $S_n$,  
which only depends on the conjugacy class $\rho=[\pi]$ of a permutation $\pi\in S_n$, given by the cycle structure of $\pi$. The conjugacy class $\rho\vdash n$ is also labeled by an integer partition.
Some examples of Weingarten functions are:
\begin{align}
\Wg{3,N}(21)&=\frac{-1}{(N^2-1)(N^2-4)},&  \Wg{3,N}(1^3)&=\frac{N^2-2}{N(N^2-1)(N^2-4)}.
\end{align}
The Weingarten functions for $a-b=q\Nc$ with $q=1$ has been addressed in \cite{Zuber2016}. For $q\neq 0$, also epsilon tensors enter Eq.~(\ref{Weingarten}), which
leads to lengthy expressions. The generalization for $q>1$ will be addressed in a forthcoming publication.
Here we simply want to illustrate that we recover the strong coupling limit and the leading order gauge correction within this formalism. 

\subsection{Link Integration via Weingarten Functions}\label{sec-1-4}

The Weingarten functions are a powerful tool to address gauge corrections and integrals for many flavors, given that the matrices $\M$, $\Md$ are generalized to $\Nf>1$.
Depending on how many fermion hoppings contribute to the link integral, we restrict the sum over irreps within the Weingarten function 
to those consistent with the fermion content:
\begin{align}
\Wg{n,N}_\lambda(\rho)&=\frac{1}{(n!)^2}\frac{(f^\lambda)^2}{D_\lambda}\chi_\lambda^\rho,
&\Wg{n,N}_\Lambda(\rho)&=\sum_{\substack{\lambda\vdash n\\\lambda\in \Lambda}}\Wg{n,N}_\lambda(\rho)
\end{align}
This restriction is possible due to the orthogonality of characters: for any $\lambda \neq [n]$ (i.e.~with the exception of the completely symmetric irrep which has $\chi_{[n]}^\rho=1$ for all 
$\rho$) it holds that
\begin{align}
\sum_{\rho\vdash n} h_\rho \chi_\lambda^\rho &=0,& \sum_{\rho\vdash n} h_\rho = n!
\end{align}
with $h_\rho$ the number of elements in the conjugacy class $\rho$. 
However, due to the additional minus signs from the ordering of the Grassmann variables, there are other irreps $\lambda\in\Lambda$ which are non-zero.  
At strong coupling, where all sources are fermionic, only the completely anti-symmetric irrep is non-zero, $\Lambda=\{[1^n]\}$, with  $n\leq \Nc$.
Here, $\chi_{[1^n]}^\rho=\sgn(\rho)$, resulting in
\begin{align}
\I^{n,n}\ijkl(\Lambda)&=\sum_{\sigma,\tau\in S_n}\prod_{r=1}^n
\left(
\delta_{i_{{\sigma(r)}}}^{\;l_{r}}
\delta_{k_{r}}^{\;j_{\tau(r)}}
\right)
\frac{1}{(n!)^2} \frac{(\Nc-n)!}{\Nc!}\sgn(\rho),
& \rho&=[\sigma\tau^{-1}]
\\
%&=\sum_{\rho\vdash n}\prod_{r=1}^n
%\tr_\rho[\ldots]
%\frac{(\Nc-k)!}{\Nc!k!}\sgn(\rho),\qquad \rho=[\sigma\tau^{-1}]\\
J_0(\M,\Md)%&=\sum_{k=0}^{\Nc}\frac{1}{(k!)^2}\sum_{\rho\vdash k}h_\rho \tr_\rho[\M\Md]\frac{(\Nc-k)!}{\Nc!}\sgn(\rho),\nonumber\\
&=\sum_{k=0}^{\Nc}\sum_{\rho\vdash k}
h_\rho \tr_\rho[\M\Md]
  \frac{(\Nc-k)!}{\Nc!(k!)^2}\sgn(\rho),\\
\tr_\rho[\ldots]&=\prod\limits_i\tr[(\ldots)^i]^{\rho_i},\qquad \sum\limits_i i\rho_i=n.
%\I^{\Nc,0}\ijkl(\Lambda)&=
%\sum_{\EE{\Nc} =1}^{\Nc}
%[\epsilon_{i_{e_1},\ldots i_{e_{\Nc}}}][\epsilon^{j_{e_1},\ldots j_{e_{\Nc}}}]
\end{align}
This agrees for $\Nf=1$ with the result in Eq.~(\ref{JSC}) since $\sgn(\rho)$ is canceled by the anti-commutativity of the Grassmann variables.

For the leading order gauge corrections, the additional gauge link from the plaquette allows partial symmetrization, $\Lambda=\{[1^n],[21^{n-2}]\}$:
\begin{align}
\I^{n,n}\ijkl(\Lambda)%=
%\sum_{\sigma,\tau\in S_n}\prod_{r=1}^n
%\left(
%\delta_{i_{d_{\sigma(r)}}}^{\;l_{r}}
%\delta_{k_{r}}^{\;j_{d_\tau(r)}}
%\right){\Wgl{n,\Nc}{[1^n]}(\rho)
%+\Wgl{n,\Nc}{[21^{n-2}]}(\rho)
%}\\
&=
\sum_{\sigma,\tau\in S_n}\prod_{r=1}^n
\left(
\delta_{i_{{\sigma(r)}}}^{\;l_{r}}
\delta_{k_{r}}^{\;j_{\tau(r)}}
\right){
\frac{1}{(n!)^2}
\frac{(\Nc-n)!}{(\Nc+1)!}\left((\Nc+1)\sgn(\rho)+(\Nc+1-n)\chi_{[21^{n-2}]}^{\rho}\right)
},
\\
%&=\sum_{\rho\vdash n}\tr_\rho[\ldots]{
%\frac{(\Nc-n)!}{\Nc!n!}\left(\sgn{\rho}+\frac{\Nc+1-n}{\Nc+1}\chi_{[21^{n-2}]}^{\rho}\right)
%}\\
%&=\sum_{\rho\vdash n}\prod_{r=1}^n
%\left(
%\delta_{i_{d_{\sigma(r)}}}^{\;l_{r}}
%\delta_{k_{r}}^{\;j_{d_\tau(r)}}
%\right){
%	\frac{(\Nc-n)!}{\Nc!(n-1)!}}\\
J\IJZZ(\M,\Md)&=\sum_{k=0}^{\Nc}\frac{1}{(k-1)!k!}\sum_{\rho\vdash k}h_\rho
\tr_\rho[\M\Md \M\IJZZ]\nonumber\\
&\qquad\quad \frac{(\Nc-k)!k!}{(\Nc+1)!}\left((\Nc+1)\sgn(\rho)+(\Nc+1-n)\chi_{[21^{n-2}]}^{\rho}\right)
%&=\sum_{k=0}^{\Nc}\sum_{\rho\vdash k}
%h_\rho\tr_\rho[\M\Md \M\IJZZ]\nonumber\\
%&\frac{(\Nc-k)!}{(\Nc+1)!(k-1)!}\left((\Nc+1)\sgn(\rho)+(\Nc+1-n)\chi_{[21^{n-2}]}^{\rho}\right)
\end{align}
For $\Nf=1$ this reproduces the known result \cite{Eriksson1981,deForcrand:2014tha}:
\begin{align}
J\IJZZ(\M,\Md)=\sum_{k=0}^{\Nc}\frac{(\Nc-k)!}{\Nc!(k-1)!}(M_xM_{x+\hmu})^k\M\IJZZ
\end{align}
With this result, one address gauge corrections as shown in Fig.~\ref{fig-2-1}.
Similarly, other gauge corrections can be addressed, which we plan to do in a forthcoming publication.

\begin{figure}[thb] % no figure before 1st section
  \centering
  \includegraphics[width=0.77\textwidth,clip]{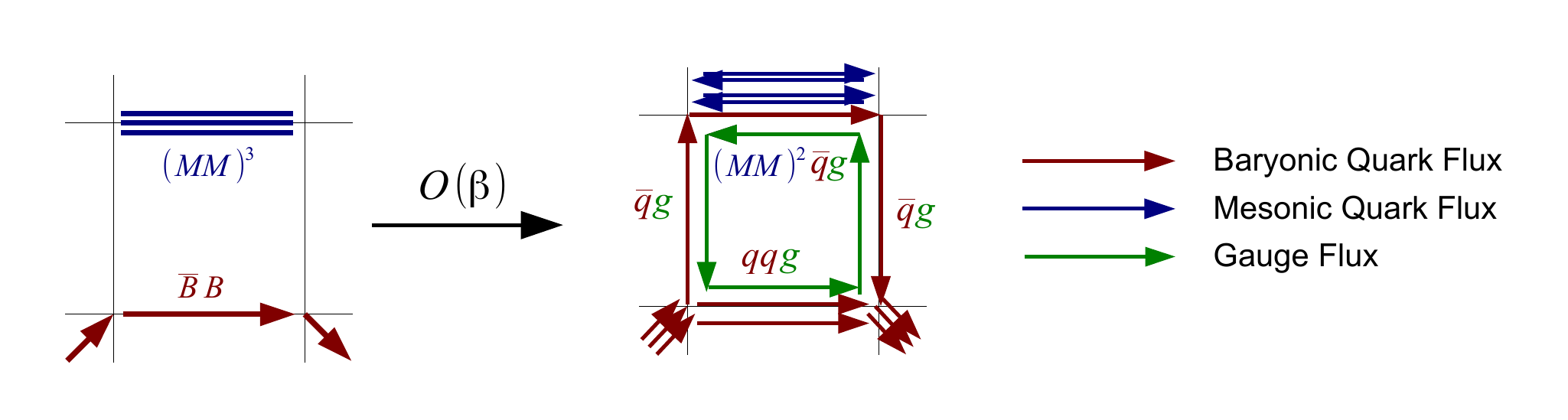}
  \includegraphics[width=0.77\textwidth,clip]{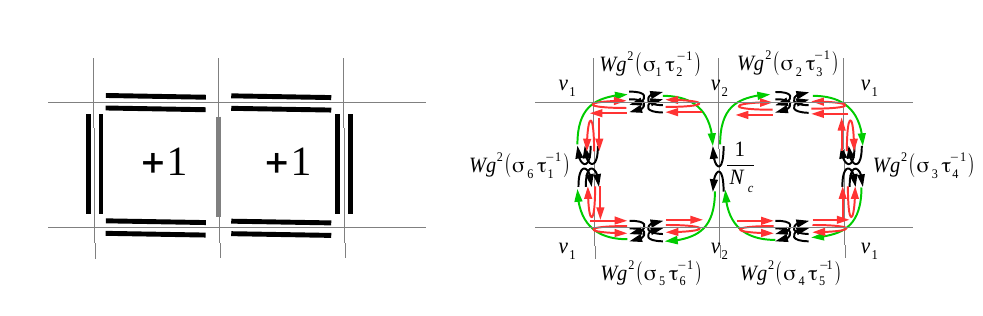}
  \caption{Gauge corrections to the strong coupling limit. Top: the effect of gauge corrections to world lines: a baryon along an excited plaquette, 
  smearing the previously point-like baryons over a lattice spacing.
  Bottom: Two excited adjacent plaquettes, displayed as dimer covering, and with internal structure of dimers. 
  Contributions from plaquettes, dimers via Weingarten functions, 
  green: plaquette contributions, red: fermion hoppings, black: permutations that enter the Weingarten functions and are summed over. 
  The vertex weights $v_1=v_2=1$ are trivial in that example.}
  \label{fig-1}% Give a unique label
\end{figure}

\section{Dual Formulation}\label{sec-2}

\subsection{Grassmann Integration}\label{sec-2-1}

\begin{figure}[thb] % no figure before 1st section
  \centering
  \includegraphics[width=0.77\textwidth,clip]{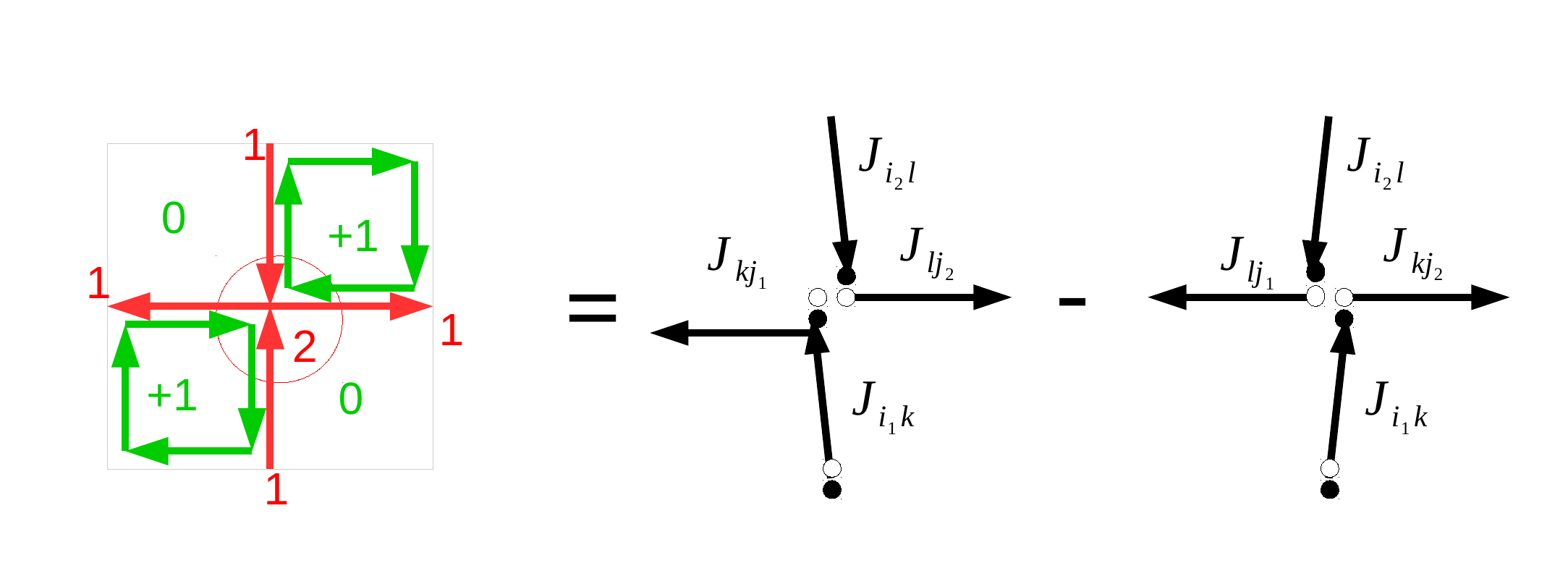}
  \includegraphics[width=0.77\textwidth,clip]{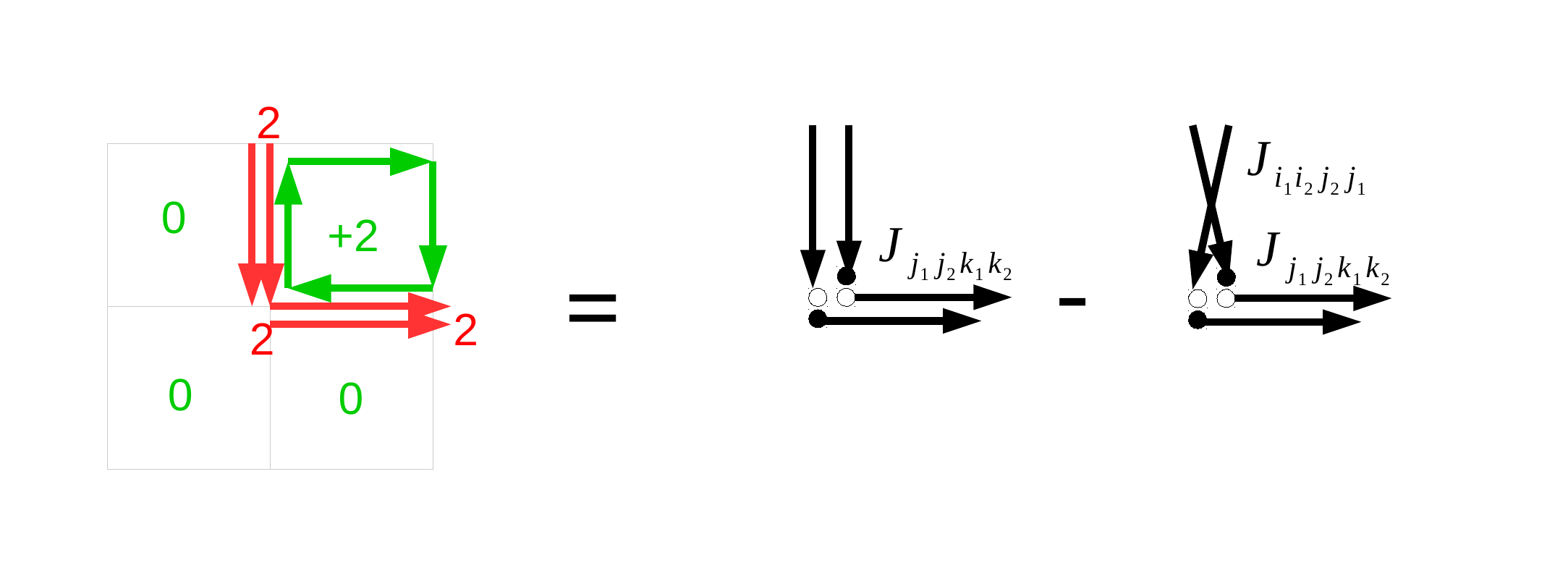}
  \caption{Simplification due to Grassmann integration within the $\U(\Nc)$ sector due to exact cancellations. Top: plaquette configurations that result in $f_x>1$.
  Bottom: plaquette configurations that result in $f_b>1$. Hence, in $\U(\Nc)$, the quark fluxes around the plaquette surfaces form self-avoiding loops.
  We will also restrict to that in $\SU(\Nc)$, where this simplification is no longer applicable and introduces systematic errors at $\mathcal{O}(\beta^\Nc)$.}
  \label{fig-2-1}% Give a unique label
\end{figure}

Given that all link integrals $\K\ijkl$ are computed, the remaining task is to organize the fermions such that they can be integrated out. If the integrals $\K\ijkl$ have more than two open indices,
the Grassmann integration gives rise to a tensor network that is difficult to evaluate. For $\U(\Nc)$ gauge theory, the contractions are however possible as there are exact cancellations, as shown in Fig.~\ref{fig-2-1}.
Only integrals with two open indices $\K^{1,0}_{i_{1}^{\;j_1}}$ or $\K^{0,1}_{k_{1}^{\;l_1}}$ give non-zero contributions. Since Grassmann integration results in one incoming and one outgoing loop per site 
if the site is on the boundary of a plaquette surface, these have to be contracted along loops.
The resulting simplifying constraint, exact for $\U(\Nc)$ and valid for the $q=0$ sector of $\SU(\Nc)$, is that plaquette surfaces are bound by quarks which form self-avoiding loops. However, for $q\neq0$, 
quark loops can intersect such that the constraint is no longer valid. We will nevertheless apply this constraint, resulting in systematic errors for fermionic observables at $\mathcal{O}(\beta^
\Nc)$.

\subsection{The Partition Function}\label{sec-2-2}

With the above simplification, the resulting partition sum is a sum over monomers, dimers, world lines and world sheets defined as surfaces of 
constant plaquette occupation numbers.
To do so, we have to introduce two auxiliary variables which are completely determined by the plaquette configuration:
\begin{align}
f_b&= \sum_{P\in P_b^+} (n_P-\bar{n}_P)+\sum_{P\in P_b^-} (\bar{n}_P-n_P)\in \{0,\pm 1\},\qquad  f_x=\frac{1}{2}\sum_b |f_b|\in\{0,1\}, \\
\ell_f &= \{ b=(x,\mu) \,| \, f_b=\pm 1 \, \, \text{are connected}  \}  \equiv \partial \{n_P,\bar{n}_P\}, 
\end{align}
where $f_b$ counts the number of fermion fluxes through a bond $b$, $f_x$ counts the number of fermion fluxes through a site $x$,
and $\ell_f$ are the self-avoiding loops that are defined on the boundary of the plaquette surfaces of constant plaquette occupation numbers.
With this, the partition function reads
\begin{align}
Z(am_q,a_t\mu,\gamma)&=\sum_{\{k_b,n_x,\ell_\Nc,n_P,\bar{n}_P\}}
%\underbrace{\prod_{b=(x,\mu)}\frac{(\Nc-k_b)!}{\Nc!(k_b-{|f_b|})!}}_{\text{singlet hoppings}\,M_x M_y}
%\underbrace{\prod_{x}\frac{\Nc!}{n_x!}(2am_q)^{n_x}}_{\bar{\psi}\psi}
%\underbrace{\prod_{\ell_3} w(\ell_3,\mu)}_{\text{triplet hoppings}\,\bar{B}_xB_y}
%\underbrace{\prod_{\ell_f} {\tilde{w}({\ell_f},\mu)}}_{\text{weight modifications}}
%\underbrace{\prod_P\frac{\left(\frac{\beta}{2\Nc}\right)^{n_P+\bar{n}_P}}{n_P!\bar{n}_P!}}_{\text{gluon propagation}}
\prod_{b=(x,\mu)}\frac{(\Nc-k_b)!}{\Nc!(k_b-{|f_b|})!}\gamma^{(2k_b-f_b)\delta_{\mu 0}}
\prod_{x}\frac{\Nc!}{n_x!}(2am_q)^{n_x}\nonumber\\	
&\hspace{2.3cm}\times\prod_{\ell_\Nc,\ell_f} w(\ell_\Nc,\ell_f,\mu)
\prod_P\frac{\left(\frac{\beta}{2\Nc}\right)^{n_P+\bar{n}_P}}{n_P!\bar{n}_P!}\\
k_b\in \{0,\ldots \Nc\},&\qquad  n_x \in \{0,\ldots \Nc\},\qquad \ell_\Nc \in \{0,\pm 1\},\qquad n_P, \bar{n}_P\in \Nat.
\label{ZMDPP}
\end{align}
Due to restriction discussed Sec.~\ref{sec-2-1}, we however only sample plaquette surfaces where either $\bar n_P=0$ or $n_P=0$, resulting in 
a net plaquette occupation number $n_P-n_P=0 \in \Zat$.

The color constraint, a modification of the Grassmann constraint, is
\begin{align}
n_x+\sum_{\hat{\mu}=\pm\hat{0},\ldots \pm \hat{d}}\lr{k_{\hat{\mu}}(x) + \frac{\Nc}{2} |\ell_{\Nc,\hmu}(x)|} =\Nc+{f_x}.
\end{align}
The $\Nc$-flux loops $\ell_\Nc$ have the same role as baryon loops at strong coupling, but they are now not necessarily made up of $\Nc$ quarks.
Likewise, also dimers are not necessarily mesons, but can be composed of a quark-gluon combination.
The bond weights are modified in case a bond is both part of a loop $\ell_\Nc$ and a loop $\ell_f$ :
\begin{align}
w(B_1)&=\frac{1}{N_c!(N_c-1)!},& w(B_2)&=\frac{{ (N_c-1)!}}{N_c!}
\end{align}
with $B_1$ a $\Nc$-flux bond without and $B_2$ with an additional dimer. 
Also the site weights are modified in case fermion flux is reoriented, i.e.~when $f_x=1$, 
with $v_1=(\Nc-1)!$ the weight when it merges into a dimer, and $v_2=\Nc!$ when it merges with a $\Nc$-flux.
\begin{figure}[thb] % no figure before 1st section
  \centering
  \includegraphics[width=0.495\textwidth,,clip]{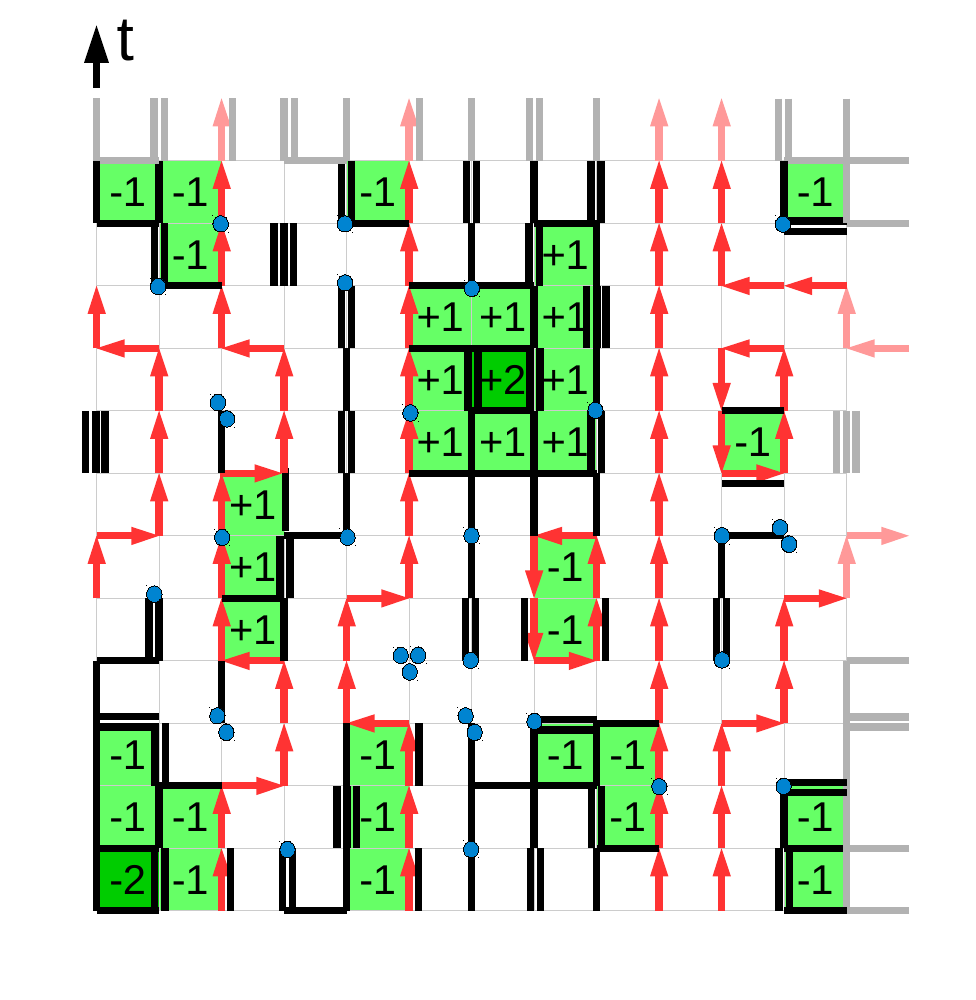}
  \includegraphics[width=0.495\textwidth,clip]{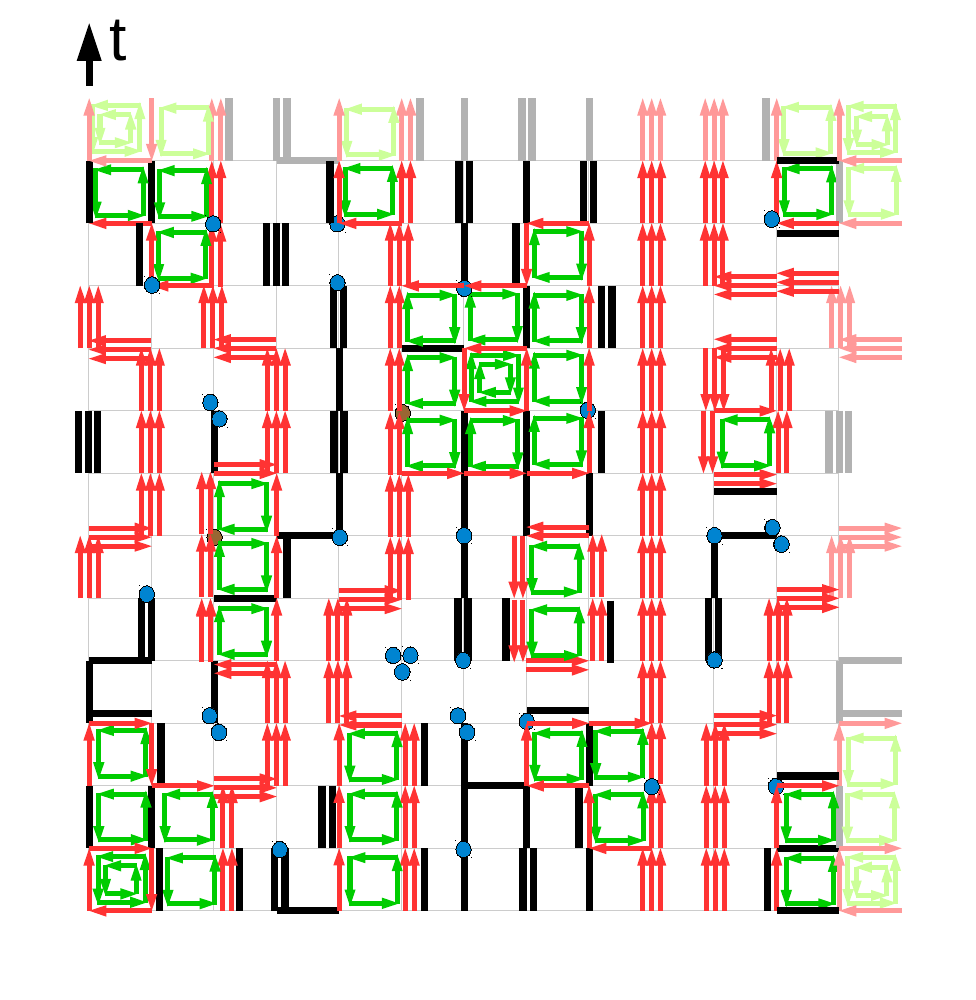}
  \caption{Typical 2-dimensional configuration at finte $\beta$, $a_t\mu$ and $am_q$. Left: degrees of freedom that are sampled: monomers (blue), dimers (black), 3-fluxes (red) and plaquette occupation numbers (green).
  Right: the same configuration but with the substructure of color singlets and triplets along excited plaquettes: quarks (red) and gauge fluxes (green). 
  Baryons becomes extended objects.
  }
  \label{fig-2-2-1}% Give a unique label
\end{figure}

We sample the partition function Eq.~(\ref{ZMDPP}) by extending the mesonic and baryonic worm algorithm used at strong coupling. 
In particular, we update the plaquette occupation numbers on closed loop configurations, and the 0-flux and $\Nc$-flux worms take modified weights on edges with $f_b\neq 0$.
A detailed discussion of the algorithm will be left for a forthcoming publication.

\subsection{Sign Problem}\label{sec-2-3}

Although the finite density sign problem has been made very mild in the strong coupling limit, this is not necessarily the case away from the strong coupling limit, 
as fermion hoppings on the boundary of plaquette surfaces take place. Single fermion hoppings are however not suppressed by a large mass.
In fact, the sign problem in the dual representation due to finite $\beta$ even arises for the $\U(\Nc)$ gauge theory,
which is sign problem-free in the conventional fermion determinant representation, as the depenence on the chemical potential drops out. 

\begin{figure}[thb] % no figure before 1st section
  \centering
  \includegraphics[width=12cm,clip]{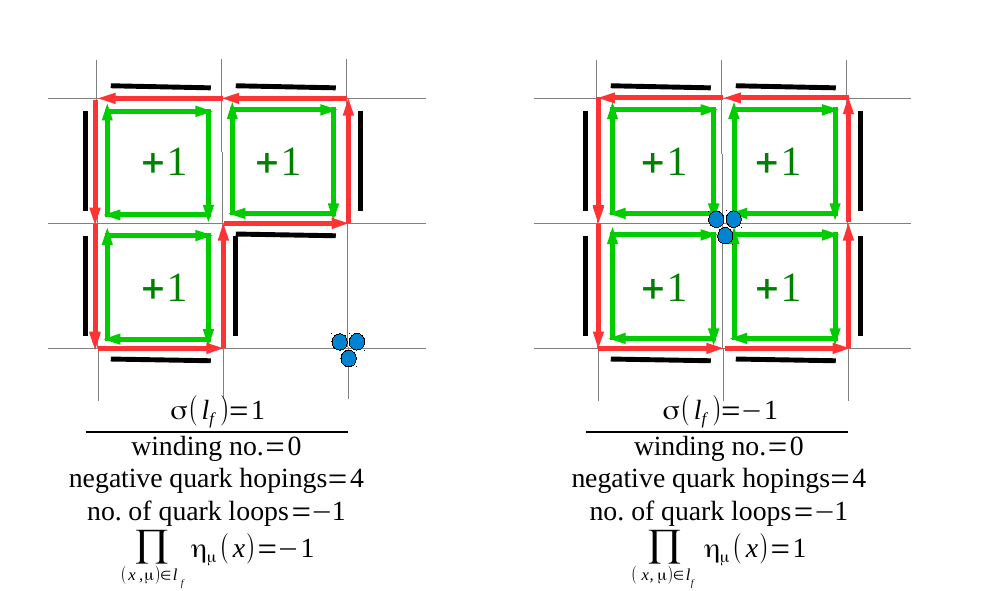}
  \caption{The plaquette-induced sign problem: example of two configurations with opposite signs.}
  \label{fig-3-1}% Give a unique label
\end{figure}

The sign of a configuration factorizes in the $\Nc$-flux sign and the fermion flux sign:
% \begin{itemize}
% \item triplet flux {\color{darkgreen}$\ell_3$} form self-avoiding loops
% \item gauge flux \tbfb{$\ell_f$} form self-avoiding loops (no winding)
%\end{itemize}
\begin{align}
\sigma(C)&=\prod_{\ell_f}\sigma(\ell_f)\prod_{\ell_\Nc}\sigma(\ell_\Nc),
&  \sigma(\ell)&=(-1)^{1+w(\ell)+N_{-}{(\ell)}}\prod_{\tilde{\ell}}\eta_\mu(x).
\end{align}
%\end{minipage}\quad
%\begin{minipage}{0.4\textwidth}
%\vspace{2mm}
%\includegraphics[width=\textwidth]{../../../Configs/config_Nt8G.pdf}\\[-6mm]
%\centerline{\footnotesize{ensemble with plaquettes}}
%\vspace{-2mm}
%\end{minipage}

For $\Nc=3$, the combination of fermion loops and 3-flux loops lead to the following identification, as shown in Fig.~\ref{fig-2-2-1}: dimers on bonds with fermion $f_b\neq 0$ are fermionic, whereas
3-fluxes on bonds with fermion $f_b\neq 0$ are bosonic.

The example of a negative configuration, Fig.~\ref{fig-3-1} (right), illustrates that in two dimensions, negative contributions are related to frustration of monomers: a loop trapping an odd number of monomers 
has negative sign. This is known from the dual representation of the Schwinger model at finite quark mass. But for dimensions $d>2$, even without monomers, a sign problem is induced as 
dimers and $\Nc$-fluxes can be perpendicular on a plaquette surface, giving rise to topologically inequivalent configurations with opposite signs.

\subsection{Crosschecks}\label{sec-2-4}

\begin{figure}[thb] % no figure before 1st section
  \centering
  \includegraphics[width=14cm,clip]{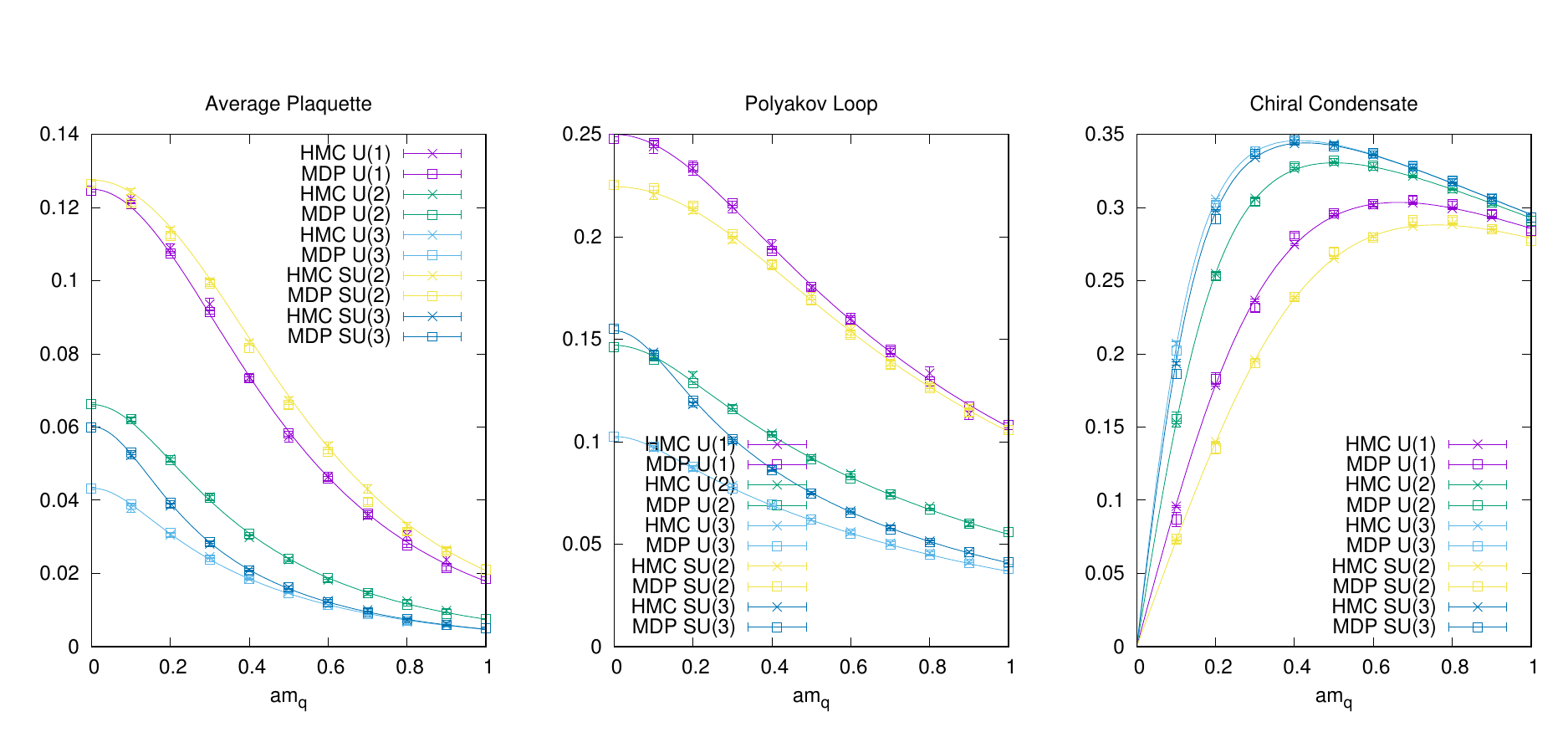}
  \caption{Crosschecks for various gauge groups at $\mu_B=0$ on small lattices where both analytic results from exact enumeration and hybrid Monte Carlo data were obtained.
  The average plaquette, Polyakov loop and chiral susceptibility are shown as a function of the quark mass.}
  \label{fig-2-4-1}% Give a unique label
\end{figure}

We have made extensive crosschecks on small 2-dimensional volumes where exact enumeration is possible. In Fig.~\ref{fig-2-4-1} some gauge observables, the average plaquette and the Polyakov loop,
are shown as obtained from the dual representation, as a function of $am_q$ for $\mu=0$, and for various gauge groups. They agree well both with the exact result and with hybrid Monte Carlo (HMC).

\begin{figure}[thb] % no figure before 1st section
  \centering
  \includegraphics[width=14cm,clip]{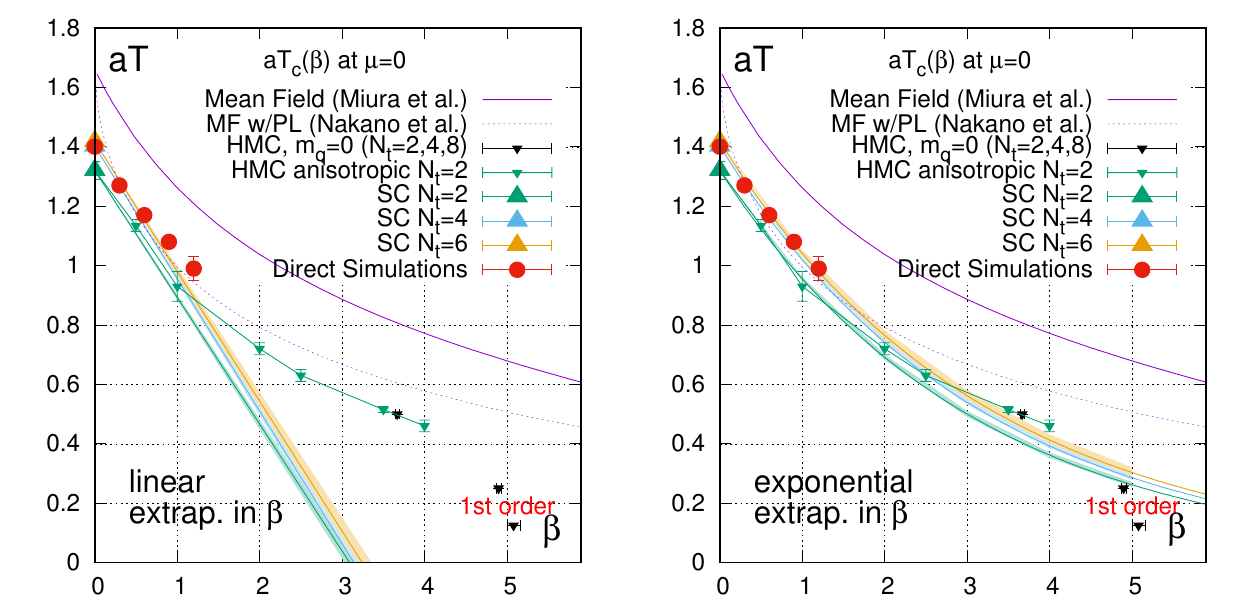}
  \caption{The phase boundary for SU(3) at $\mu_B=0$. The comparison of direct sampling (red dots) with reweighting and mean field theory. This results makes use of the mean field value of $a/a_t=\gamma^2$ for better comparison.
  The direct simulations favor the scenario of extrapolating the phase boundary via an exponential ansatz (right) rather than a linear ansatz (left), 
  as has been discussed in \cite{deForcrand:2014tha}.
  }
  \label{fig-2-4-2}% Give a unique label
\end{figure}

Another important crosscheck where HMC and Meanfield results \cite{Miura:2016kmd} are available is the phase boundary in the 
$\beta$-$T$ plane for SU(3) at $\mu=0$. Fig.~\ref{fig-2-4-2} shows that the results from direct sampling agree well with extrapolations of HMC. 

\section{Results on the Phase Diagram}\label{sec-3}

\subsection{Strong Coupling Regime at Finite Temperature}\label{sec-3-1}

We have derived the dual representation in the strong coupling limit by taking into account the bare anisotropy $\gamma$ in order to continuously vary the temperature independent of $\beta$.
In a recent publication \cite{deForcrand:2017fky}, one of us has determined with collaborators the non-perturbative anisotropy $a/a_t$ as a function of the bare anisotropy in order to unambiguously define the temperature:
\begin{align}
aT&=\frac{\xi(\gamma)}{\Nt},& \frac{a}{a_t}\equiv \xi(\gamma)\simeq \kappa+\frac{1}{1+\lambda\gamma^4},\qquad \lambda=\kappa/(1-\kappa),\qquad \kappa\simeq 0.7810(8) 
\end{align}
We adopt this non-perturbative definition of the temperature, which differs significantly from the previously used mean field result $aT=\frac{\gamma^2}{N_t}$.
Likewise we convert the chemical potential: $a\mu_B=\xi(\gamma) a_t \mu_B$.

\subsection{Phase Diagram in the Strong Coupling Regime}\label{sec-3-2}

\begin{figure}[thb] % no figure before 1st section
  \centering
  \includegraphics[width=12cm,clip]{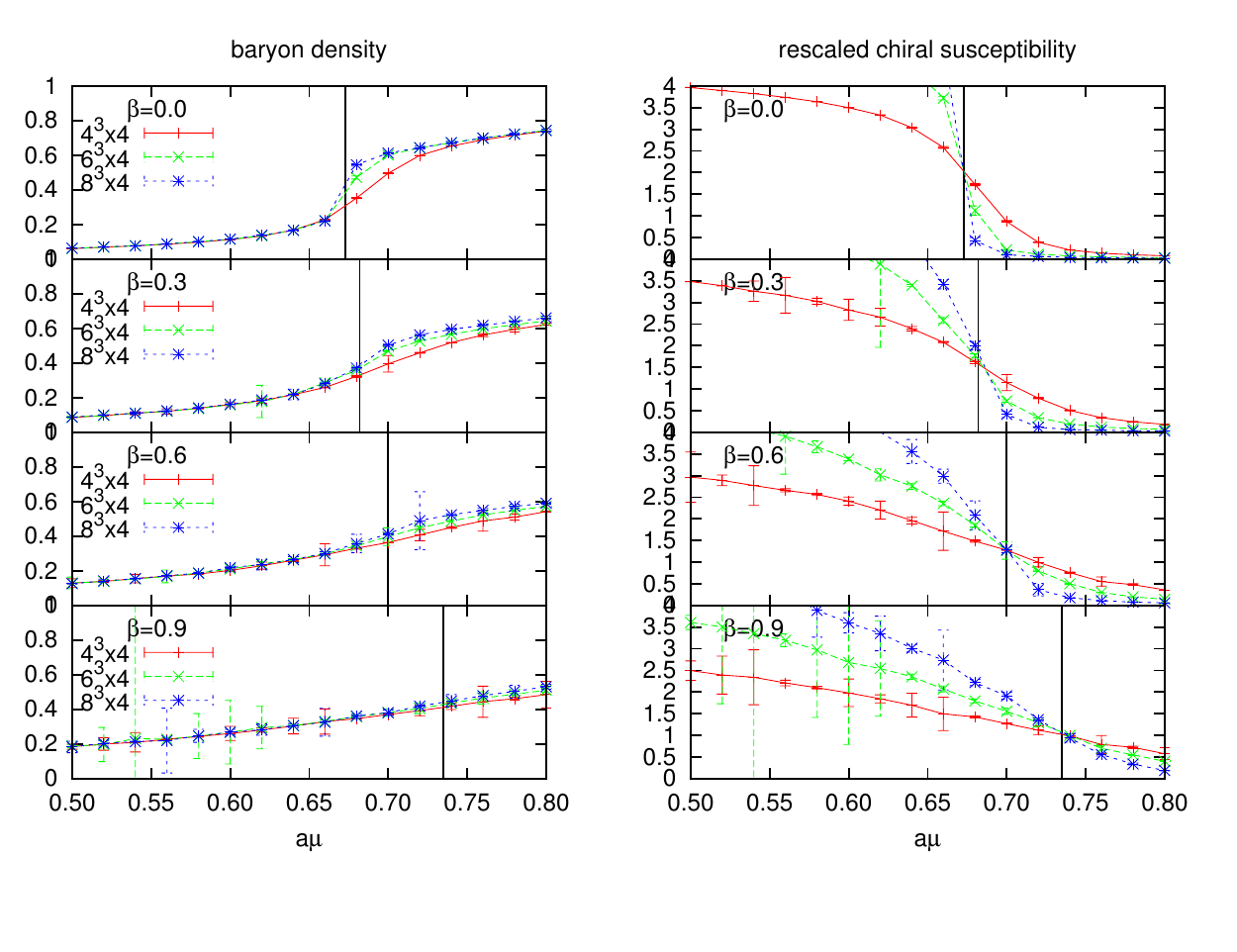}
  \includegraphics[width=12cm,clip]{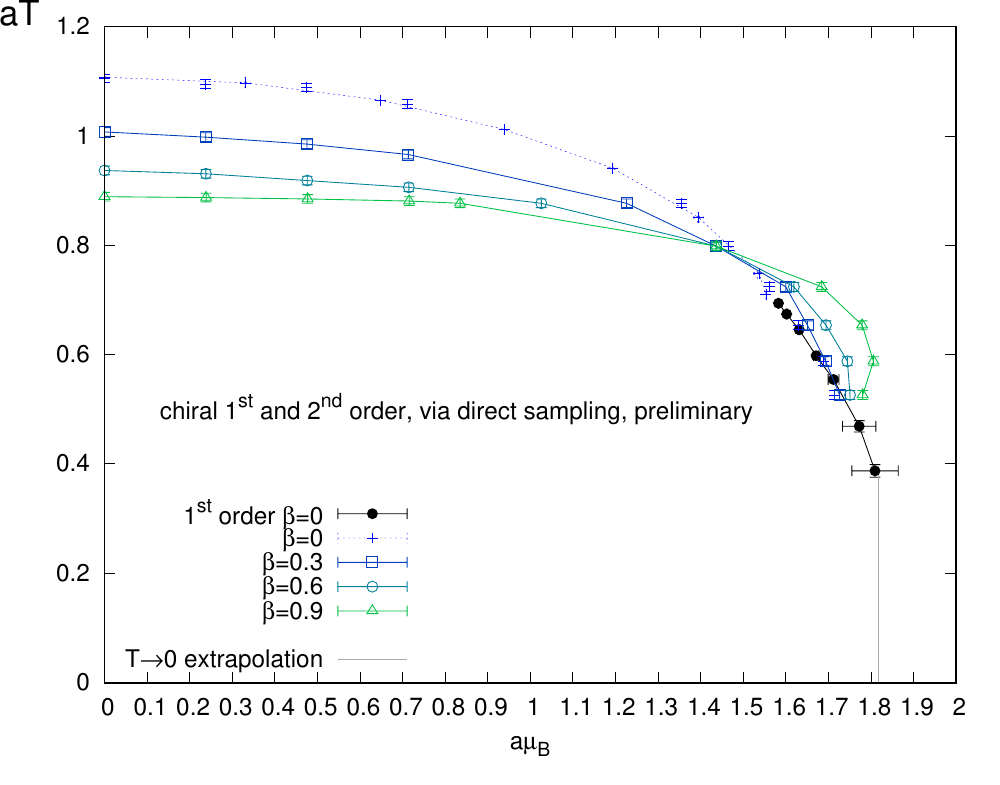}
  \caption{Top: baryon density and rescaled chiral susceptibility at $aT=0.7$ (in the vicinity of the tri-critical point) from direct simulations for various $\beta$. 
  Bottom: The phase boundary for SU(3) at $\mu_B=0$. For $\mu_B=0$ up to the tri-critcial point, 
  the direct simulations agree well with the results from reweighting, but a different behaviour is observed along the first order line.}
  \label{fig-3-2-2}% Give a unique label
\end{figure}

Lattice QCD with staggered fermions has a residual chiral symmetry even in the strong coupling regime, since there is an exact Goldstone mode in the spin$\otimes$taste basis 
$\gamma_5\otimes\gamma_5$. The lattice action at zero quark mass, and likewise partition function Eq.~(\ref{ZMDPP}) has the symmetry
\begin{align}
U(1)_V\times U(1)_{55} :&&\chi(x)\mapsto e^{i\epsilon(x)\theta_A+i\theta_V}\chi(x),\quad \epsilon(x)=(-1)^{x_1+x_2+x_3+x_4},
\end{align}
i.e.~even and odd sites transform independently. The chiral symmetry is spontaneously broken at low temperatures, but restored at some phase boundary
$aT_c(a\mu_B)$. The transition in the chiral limit is second order for small and intermediate $a\mu_B$ and turns into a first order transition at low temperatures, 
separated by a tri-critical point. This point at 
$(a\mu_B^{\rm tric} , aT^{\rm tric} ) = (1.56(4), 0.73(4))$
turns into a critical end point as soon as the quark mass becomes finite.
The ratio $\mu_B^{\rm CEP}/T^{\rm CEP}>2$ becomes even larger as a function of the quark mass.
The phase boundary for the chiral transition in the strong coupling regime can be measured by finite size scaling of the chiral susceptibility, 
as shown in Fig.~\ref{fig-3-2-2} (top). The nuclear transition can be obtained from the position of the gap in the baryon density.
In the strong coupling limit, the first order chiral and nuclear transition coincide.
The reason is that the nuclear liquid phase is actually a 
Pauli saturated phase of a baryon crystal, such that no quarks are left for the formation of a chiral condensate.
This finding seems to be independent of the quark mass \cite{Kim:2016izx}.
We restrict in the following to the chiral limit, where simulation via the Worm algorithm are even faster than with finite quark mass, in contrast to HMC.\\

\begin{figure}[thb] % no figure before 1st section
  \centering
  \includegraphics[width=0.58\textwidth,clip]{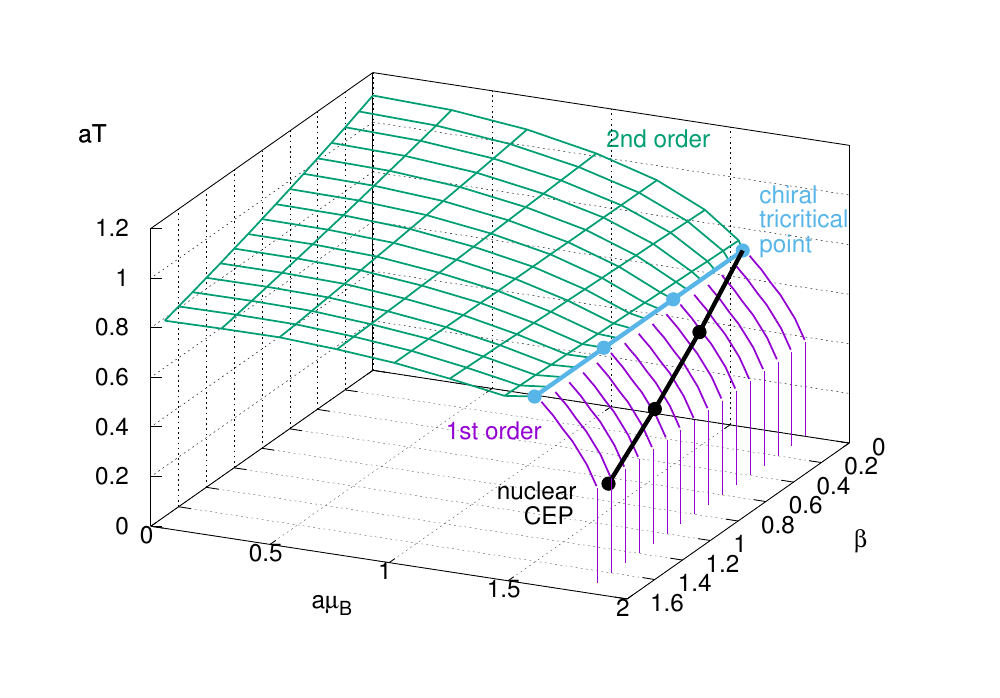}
  \includegraphics[width=0.41\textwidth,clip]{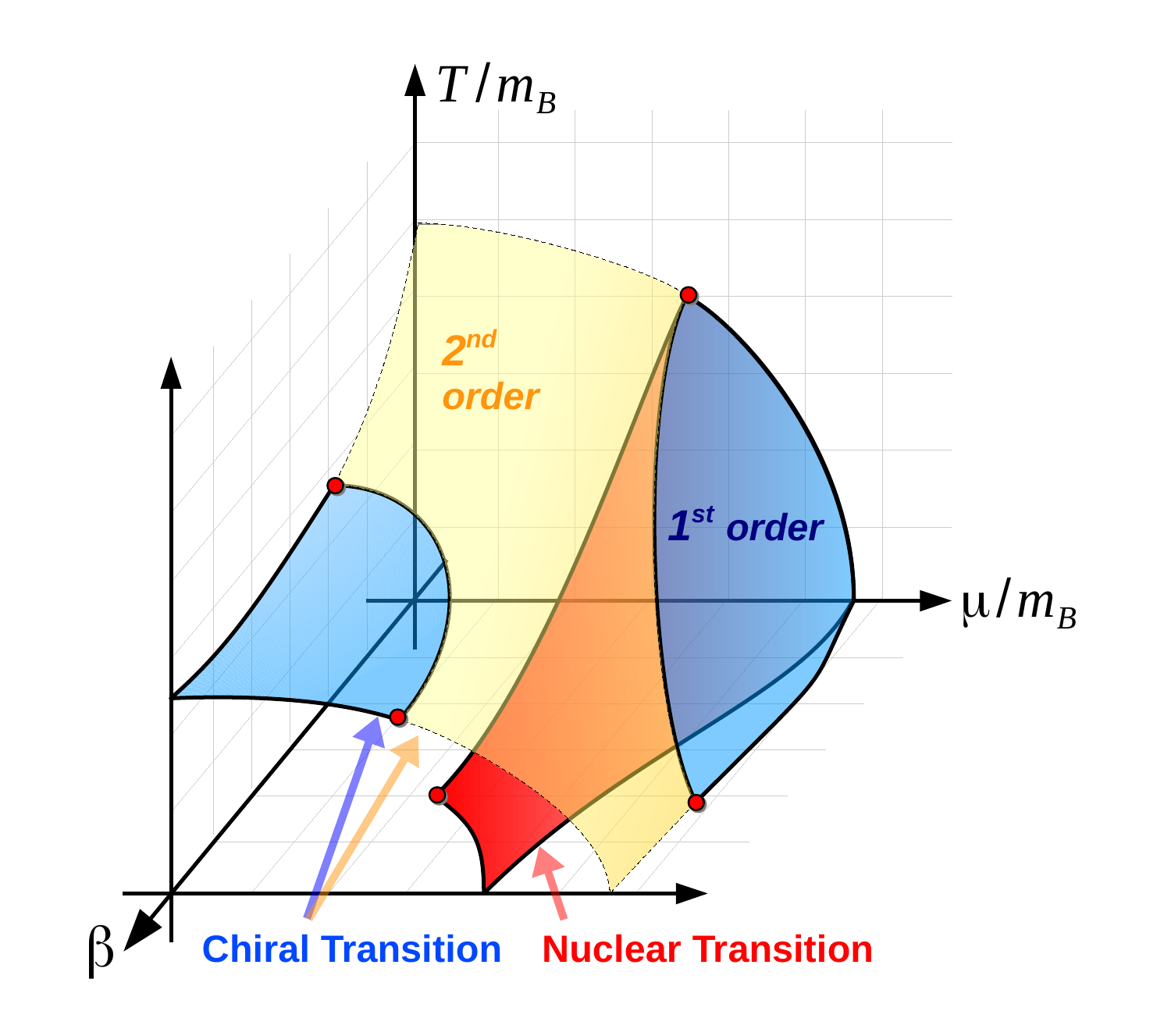}
  \caption{Left: The phase boundary for SU(3) in the chiral limit as a function of small $\beta$, obtained from reweighting \cite{deForcrand:2014tha} but with the non-perturbative anisotropy $a/a_t$ 
  to convert to $aT$ and $a\mu_B$.  
  Contrary to the expectation, the nuclear and chiral transition did not split, 
  which is likely an artifact from reweighting. Right: one of several possible scenarios on the $\beta$-dependence of the chiral and nuclear transition for unrooted staggered fermions in the chiral limit.
  }
  \label{fig-3-2-1}% Give a unique label
\end{figure}

Via reweighting from the $\beta=0$ ensemble, Fig.~\ref{fig-3-2-1} (left), 
it was found that the chiral transition $aT_c(a\mu_B)$ for small chemical potential indeed decreases, 
as expected since the lattice spacing $a(\beta)$ becomes smaller. However, the chiral and nuclear first order transition still coincide with the strong coupling result for small $\beta$.
This may be very likely a reweighting artifact, as it is impossible to reweight from one phase to another phase across a first order transition. 
We only found that the nuclear critical end point separates from the chiral tri-critical point, but does not split from the first order line.
The expectation is however that the chiral and nuclear transition split, a possible scenario is shown in Fig.~\ref{fig-3-2-1} (right). 
It is however a priori not clear how much $\mu_c^{\rm nuclear}$ and $\mu_c^{\rm chiral}$ are separated in nature, and how large $\beta$ needs to be to observe 
that splitting.

In order to understand the relation between nuclear and chiral transition, we need to sample the partition function Eq.~(\ref{ZMDPP}) directly at finite $\beta$.
With the direct simulations at finite $\beta$, based on local plaquette updates together with the worm to update the dimers and 3-flux world lines, 
we find that the chiral first order transition indeed depends on $\beta$, as shown in Fig.~\ref{fig-3-2-2} (bottom). Our lattices were 
$N_s times 4$ with $N_s=4,6,8$, and for various temperatures and baryon chemical potentials, which suffices to determine the chiral phase boundary quite accurately.
These preliminary results still needs to be reconciled with the first order nuclear transition, which requires larger volumes.

\section{Conclusion}\label{sec-4}

We have presented a partition function that includes higher order gauge corrections with the constraint that the plaquette world sheets are bound by fermion loops.
Plaquette occupation numbers are in principle unbounded, such that we sample contributions of the gauge action at arbitrarily large order in $\beta$.
However, due to the complicated non-local structure of the tensors $C(\beta,\{S_P,S^\dagger_P\})\jilk$, it is not yet possible to write down a 
partition function that is correct for all orders in $\beta$. Hence we restrict to the limit where plaquettes form surfaces bounded by quark flux. 
This restriction is no longer valid for $\SU(\Nc)$, and our approximation will result in systematic errors in fermionic observables at $\mathcal{O}(\beta^\Nc)$.
However, in the strong coupling regime with $\beta \ll 2\Nc$, these systematic errors are expected to be small.

Due to the sign problem induced by the boundaries of the plaquette surfaces, simulations are restricted to $\beta\lesssim 1$. We presented first direct measurements at non-zero $\beta$ and $\mu$, 
which are consistent with the previous results from reweighting. 
It will be essential to improve on the sign problem further to apply these methods for $\beta>1$.

A systematic error on the phase boundary as shown in Fig.~\ref{fig-3-2-2} is due to the anisotropy $\xi=\frac{a}{a_t}$.
We only considered the bare anisotropy $\gamma_F\equiv\gamma$ in the Dirac coupling, but one should also introduce an anisotropy in the Wilson action, 
$\gamma_G=\beta_t/\beta_s$. Then the lattice anisotropy is a non-perturbative function of both bare anisotropies, $\xi(\gamma_F,\gamma_G)$, 
that can in principle be determined in a similar way as in \cite{deForcrand:2017fky}. 

In this work we have only studied the gauge corrections of the phase diagram in the chiral limit. We plan to study the gauge corrections also at finite quark mass.

\subsection{Acknowledgement}\label{sec-4-1}

We would like to thank Philippe de Forcrand and H\'elvio Vairinhos for stimulating discussions.
This work is supported by the Emmy Noether Program under the grant UN 370/1-1.
Computations have been carried out on the OCuLUS cluster at PC2 (Universität Paderborn).

%----------------------------------------------------------------------------

%%%%%%%%%%%%%%%%%%%%%%%%%%%%%%%%%%%%%%%%%%%%%%%%%%%%%%%%%%%%%%%%%%%%%%%%%%%%%
\end{document}